\documentclass[aps,pre,preprint,groupedaddress,amssymb]{revtex4}
\usepackage{graphicx}
\usepackage[figuresright]{rotating}
\begin{document}

\title{On the Evolution of Time Horizons in Parallel and Grid Simulations}
\author{L.N.\ Shchur$^{1,2,a)}$ and  M.A.\ Novotny$^{1,b)}$}
\affiliation{$^{1)}$Department of Physics and Astronomy and 
ERC Center for Computational Sciences, 
Mississippi State University, 
Mississippi State, MS 39762-5167, USA \\
$^{2)}$Landau Institute for Theoretical Physics, 142432 
Chernogolovka, Russia \\
e-mail: \tt $^{a)}$lev@itp.ac.ru, $^{b)}$novotny@erc.msstate.edu}

\begin{abstract}

We analyze the evolution of the local simulation times (LST) in
Parallel Discrete Event Simulations. The new ingredients introduced
are i) we associate the LST with the nodes and not with the processing
elements, and 2) we propose to minimize the exchange of information
between different processing elements by freezing the LST on the
boundaries between processing elements for some time of processing and
then releasing them by a wide-stream memory exchange between
processing elements. Highlights of our approach are i) it keeps the
highest level of processor time utilization during the algorithm
evolution, ii) it takes a reasonable time for the memory exchange
excluding the time-consuming and complicated process of message
exchange between processors, and iii) the communication between
processors is decoupled from the calculations performed on a
processor.  The effectiveness of our algorithm grows with the number
of nodes (or threads). This algorithm should be applicable for any
parallel simulation with short-range interactions, including parallel
or grid simulations of partial differential equations.  
\end{abstract}


\maketitle

\section{Introduction}

Parallel and grid computations for simulation of
spatially-decomposable models with short-range interactions are
arguably the most important topic in traditional applied computer
science today.  This is because of their application to simulations of
many models in science, engineering, social science, manufacturing,
and economics.  We will first concentrate our discussion on parallel
discrete event simulations (PDES), and will show the relationship to
all short-ranged grid or parallel simulations in the last sections.

PDES are the execution of a single discrete event simulation program
on a parallel computer or on a cluster of computers~\cite{Fuji90}. It
is a challenging area of parallel computing and has numerous
applications in physics, computer science, economics and engineering.
The number of applications are constantly growing in areas where
extensive dynamical processes need to be simulated, especially those
requiring a huge memory or wall-clock execution time.  For such
parallel simulations the system to be simulated is broken spatially
into disjoint subsystems, and each processing element (PE)  performs
simulations on a particular subsystem.  The simulated system jumps
discontinuously from one state to another, these jumps are called an
{\em event}. Thus the changes of state occur at {\em discrete} points
in simulated time, although the time is considered to be continuous.  
The main challenge is to efficiently process with discrete events {\em
without\/} changing the order in which the events are processed, {\em
i.e.\/} preserving the causality in the simulation.  One technique to
help preserve causality is to introduce the idea of a local virtual
time \cite{Jeff85} on a node or PE, which leads to a surface of local
simulated times (LST).

For example, consider a kinetic Monte Carlo simulation for a 2d
$L$$\times$$L$ ferromagnetic 
Ising model on a square lattice.  The discrete events
are spin flips.  If the program is to be simulated using $N_{\rm PE}$
processors, each processor may be allocated an equal
spatially-disjoint sublattice of spins.  The average interval between
two flips of the same spin varies for Metropolis-like algorithms from 
about 3.3 Monte Carlo steps per
spin (MCS)  at the critical point $T_{\rm c} \approx 2.609$ to 142.9
MCS at the low temperature of $T=0.5$~\cite{Over00}.  Implementations
of such PDES for kinetic Ising models have been performed using both
conservative \cite{Korn99} and optimistic \cite{Over00} methods of
preserving causality in the simulations.  In the conservative
implementation, a PE waits until causality is not violated before it
proceeds with its calculations.  In the optimistic implementation, if
causality may be violated the calculation proceeds using some guess, 
and if this guess is incorrect the PE must roll back to an earlier
state before any causality violations were present.

In this paper we investigate the dynamics of a number of PDES schemes.
These parallel schemes are applicable to a wide range of stochastic
cellular automata with local dynamics, where the discrete events are
Poisson arrivals. We are interested in the evolution of the time
horizon formed with the local simulation time (LST) of the nodes. In
contrast with the previous work \cite{Korn00,Kola02,Kola03,Kola04},
where each PE manages only one node and communication between PEs is
implemented according to the conservative scheme~\cite{Fuji90}, we
generalize the scheme so each PE processes a number of nodes which
communicate conservatively, whereas nodes from different PEs
communicate according to either the conservative or optimistic
scenario. The simulated time horizon is analogous to a growing
surface. The local time increments of the node correspond to the
deposition of some amounts of ``material'' at the given element of the
surface and the efficiency (which is the fraction of non-idling PEs)
of the conservative scheme exactly corresponds to the density of local
minima in the surface model. It was shown in \cite{Korn00}, that the
density of the local minima does not vanish when the number of PEs
goes to infinity. This remarkable result insures that the simulated
time horizon propagates with a nonzero velocity and that the compute
phase of the algorithm is asymptotically scalable.

The width of the time horizon in conservative PDES, after an
early-time regime and before saturation, diverges with an exponent
consistent with the Kardar-Parisi-Zhang one~\cite{Korn00}. This scaling
property is valid for the averages over the ensemble of the surfaces,
or for the ensemble of the time horizons in the language of PDES. It
is informative however to analyze the dynamics of a single realization
of the time horizon as well.  A single realization is strongly
affected by the surface fluctuations, and the evaluation of a
particular surface sometimes loses the full predictability. It is not
clear how these fluctuations are connected with the turbulence of the
Burgers solitons, although there are some similarity between
LST-horizon evolution and the evolution of the surface slope described
by noisy Burgers equations~\cite{Gurbatov,Chekhlov,Fog-noisy}.

In the original paper~\cite{Korn00} each PE simulates only one node in
a conservative manner. The efficiency of this algorithm is about
$1/4$, on average, {\em i.e.\/} one PE out of four is working at any
given time.  Each PE sends messages to its neighbors about once in
four time slices. One way to increase the utilization is to have each
node contain a large portion of the lattice, then the utilization can
be increased \cite{Korn99,Kola02,Kola03,Kola04}.

We analyze here a more efficient implementation of the Korniss {\em et
al.\/} idea.  Namely, we realize each node as a thread.  Threads are
distributed among processor elements so each PE is responsible for
some number of threads. The threads within one PE communicate using
the conservative PDES manner.  Communications among threads on the
same PE do not require the inter-processor communication latency time
(one needs only system calls within a PE), while inter-processor
communication requires calls to the I/O routines.  Processes
communicate within a single PE using system calls of the operating
system.  Inter-processor communication requires some I/O operations
involved in addition to system calls.  Threads were invented just to
accelerate communications of the partially independent parts of the
program. Their communication is supported by the kernel of modern
operation systems~\cite{Threads}.

We have two choices for the inter-processor communication: either
conservative and optimistic~\cite{Fuji90}.

With the conservative inter-processor communications, the evolution of
the time horizon will be exactly the same as in~\cite{Korn00}, albeit
the utilization would be larger than $1/4$ and close to unity for
large enough $\ell$. Here $\ell$ is the number of nodes on a PE, so
the system size simulated is $L=\ell N_{\rm PE}$.  We assume that the
check of the local minima condition (to avoid causality violations)
between threads (nodes within a PE)  is negligibly small in comparison
with the time needed to process the event.

With optimistic inter-processor communications, the evolution of the
time horizon becomes more complex. The optimistic
scenario~\cite{Fuji90} assumes that messages were not sent during some
given time window. Causality is then checked. In the case where some
node proceeds with broken causality, anti-messages have to be sent
between processors, the corresponding events are rejected.  This leads
to a roll-back to an earlier time and state.

The possible scenarios can be understood taking into account the
mapping of our problem onto surface growth dynamics. The messages sent
by one PE to another are nothing but the boundary conditions in our
algorithm. In fact, there are three (and not the two!)  possible
boundary conditions for the most left and the most right nodes for the
chain of nodes within a PE: continuous, free and fixed. Continuous
boundary conditions imply that the boundary nodes would follow the
causality, taking messages from the neighbor node on the neighbor PE.
So, this corresponds to the totally conservative case.

A more interesting solution is with fixed boundaries.  In this case, a
slope develops at boundaries of each PE. The angle of the slope
depends on the mean of the event time intervals and a nearly flat top grows
according to the conservative rule. This solution can be interpreted
as a fixed soliton of the corresponding Burgers equation.

Free boundary conditions lead to an optimistic implementation of the
algorithm. The evolution of the time horizon with the free boundaries
have mixed features compared with the algorithms with continuous or
fixed boundaries.  This boundary condition will be analyzed elsewhere.

The algorithm we discuss here is a new PDES implementation scheme.  
The main purpose of the paper is a detailed analysis of this PDES
algorithm scheme with fixed boundary conditions. The paper is
organized as follows. In section \ref{PDES-KPZ} we review briefly the
main ideas of PDES and the approach of Korniss {\em et al.\/}
\cite{Korn00}. In section \ref{Froze} we introduce our algorithm and
discuss its behavior for simple realizations in one dimension, two
dimensions, and for the solution of partial differential equations. We
discuss in \ref{Disc} the results and provide a more general overview.

\section{Conservative PDES}
\label{PDES-KPZ}

The PDES method is a tool capable of parallelizing any discrete event
dynamic algorithm, even those which appear to be intrinsically
sequential ones.  One example is the development of a kinetic Ising
model algorithm by Lubachevsky~\cite{Luba88}, and its successful
implementation~\cite{Korn99} which preserves the original dynamics of
the model.

\subsection{PDES and Kardar-Parisi-Zhang equation}

Korniss {\it et al.\/}~\cite{Korn00} developed an approach for the
analysis of such algorithms.  They mainly considered the case of
one-dimensional systems with only nearest-neighbor interactions and
periodic boundary conditions. (See ref.~\cite{KornL00} for two and
three-dimensional cases.)  Each site of the Ising model is associated
with one PE.  Consequently the original model has $N_{\rm PE}=L$ PEs
simulating $L$ node (or site).  
The number of nodes per PE are $\ell=1$.  
Update attempts at each node are independent Poisson processes with
the same rate.  (In an actual simulation the rate for the kinetic
Monte Carlo for the Ising model depends only on the energy change for a
single spin flip.)  At each PE, the random time interval $\eta_i$
between two successive attempts is exponentially distributed.

Let us associate with PE number $i$ the value of a local simulated
time $\tau_i(t)$, where we denote by $t$ the discrete time of the
parallel steps simultaneously performed by each PE.  We start with
zero LST on all PEs, {\it i.e.\/} $\tau_i(0)=0$, $i=1,2,\ldots,L$. For
$t\ge 1$ the LST evolves iteratively as \begin{eqnarray}
\tau_i(t+1)=\tau_i(t)+\eta_i(t) & {\rm if} & \tau_i(t)\le {\rm min}
\left\{ \tau_{i-1}(t),\tau_{i+1}(t) \right\} ; \nonumber \\
\tau_i(t+1)=\tau_i(t) & {\rm else} & , \label{horizon-eval}
\end{eqnarray} where $\eta_i$ are random exponential variables. Each
time the PE number $i$ advances in time it sends messages to the right
$(i+1)$ and left $(i-1)$ PEs with the time stamp of it's LST
$\tau_i(t)$. This insures that the updating process
(\ref{horizon-eval}) does not violate causality. This category of PDES
is called the {\em conservative} PDES scenario.

The Korniss {\it et al.\/} algorithm is free of deadlock, since at
least the PE with the absolute minimum LST can proceed. The efficiency
of the algorithm is the fraction of nonidling PEs and exactly
corresponds to the density of local minima of the simulated time
horizon.

The iterative process (\ref{horizon-eval}) can be rewritten as
\begin{eqnarray}
\tau_i(t+1)=&&\tau_i(t)+\Theta\left(\tau_{i-1}(t)-\tau_i(t)\right) 
\nonumber \\
 && \times \Theta\left(\tau_{i+1}(t)-\tau_i(t)\right)\eta_i(t)
\label{horizon-Theta}
\end{eqnarray}
using the Heaviside step function $\Theta$.

Introducing the local slopes $\phi_i=\tau_i-\tau_{i-1}$, the density of 
local minima can be written as 
\begin{equation}
u(t)=\frac{1}{L} \sum_{i=1}^L \Theta\left(-\phi_i(t) \right) 
\Theta\left(-\phi_{i+1}(t) \right)
\label{u-density}
\end{equation}
and its average 
\begin{equation}
\left\langle u(t) \right\rangle = 
\left\langle \Theta\left(-\phi_i(t)  \right)  
\Theta\left(\phi_{i+1}(t) \right) \right\rangle 
\label{u-speed}
\end{equation}
is the mean velocity of the time horizon, equal to 
$0.246410(7)$. Hence, the efficiency of the algorithm (in this 
worst-case scenareo) is about 25 per cent.  

It was argued by Korniss {\it et al\/}~\cite{Korn00} 
that the coarse-grained slope
of time horizon $\hat{\phi}(x,\hat{t})$ in the continuum limit evaluates
according to the Burgers equation~\cite{Kard86}
\begin{equation}
\frac{\partial \hat{\phi}}{\partial \hat{t}} = \frac{\partial^2 
\hat{\phi}}{\partial x^2} - \lambda 
\frac{\partial \hat{\phi}^2}{\partial x}
\label{Burgers-cg}
\end{equation}
and the coarse-grained time-horizon $\hat{\tau}$, $\left( \hat{\phi}=\partial
\hat{t}/\partial x \right)$ obeys the Kardar-Parisi-Zhang (KPZ) equation
\begin{equation}
\frac{\partial \hat{\tau}}{\partial \hat{t}} = \frac{\partial^2
\hat{\tau}}{\partial x^2} - \lambda \left(
\frac{\partial \hat{\tau}}{\partial x}\right)^2
\> ,
\label{KPZ-cg}
\end{equation}
which should be extended with the noise to capture the
fluctuations. 

\subsection{Time horizon evaluation in conservative PDES}
\label{THorz}

The Monte Carlo simulations of the process (\ref{horizon-eval})
showed~\cite{Korn00} that the average width of time horizon
\begin{equation}
\left\langle w^2(t)\right\rangle=\frac{1}{L}\left\langle\sum_{i=1}^L 
\left(\tau_i(t)-\bar{\tau}(t)\right)^2\right\rangle,
\label{width}
\end{equation}
where $\bar{\tau}(t)=(1/L)\sum\tau_i(t)$, grows for
appropriately chosen times (before the LST saturates) 
as $\left\langle w^2(t)\right\rangle\propto t^{2\beta}$ with
the exponent $\beta$ close to the KPZ exponent, $\beta$$=$$1/3$.
Running a single PDES on a parallel computer the LST horizon develops 
as a particular realization of the stochastic growth process, not as 
the average process.  

\begin{figure}
\centering
\includegraphics[angle=0,width=\columnwidth,keepaspectratio]{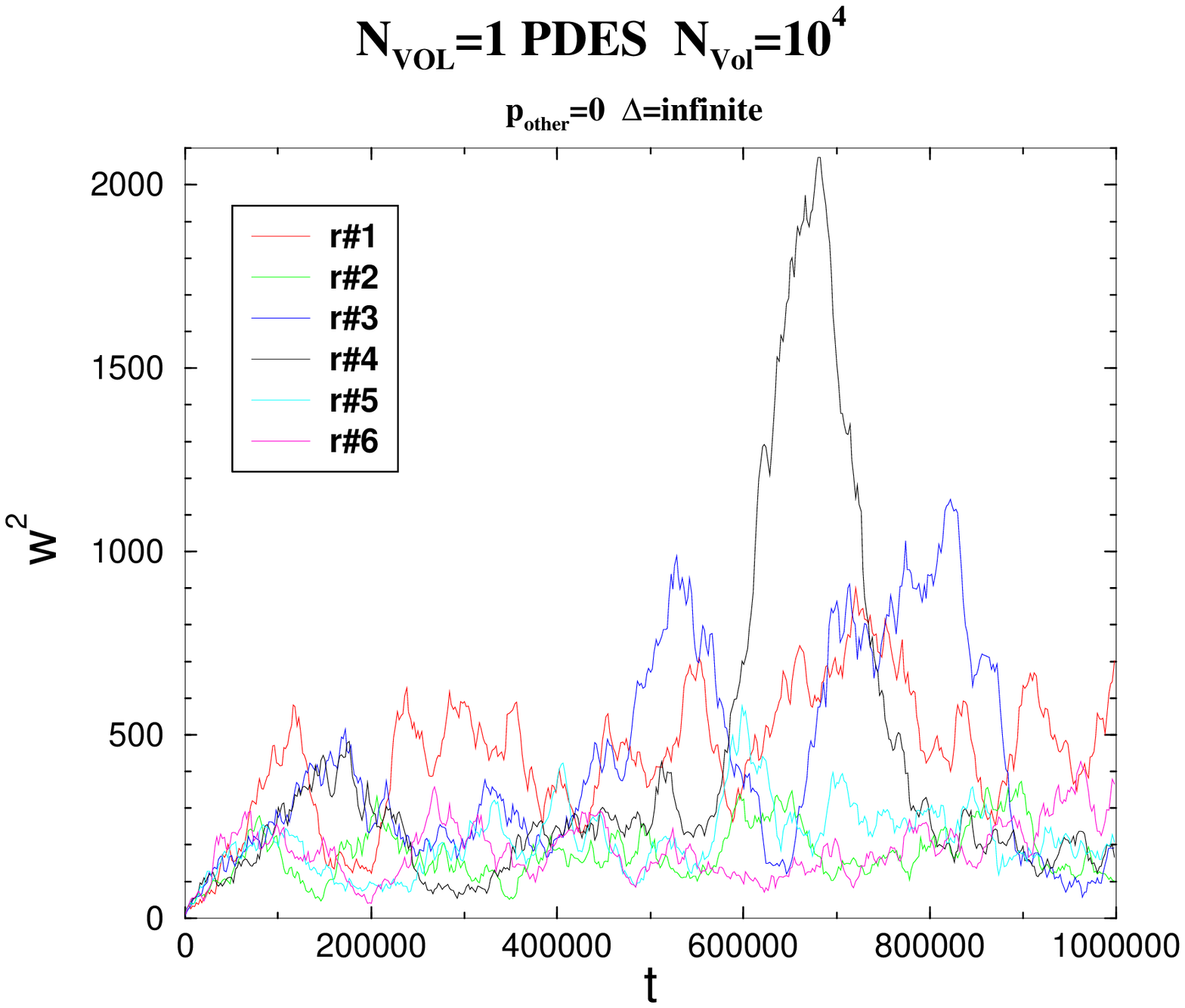}
\caption{Evolution of the width of the time horizon for several single 
realizations of conservative PDES.  Here 
there is one site per PE, and $N_{\rm PE}=L=10^4$.}
\label{NPE01000runs}
\end{figure}

\begin{figure} 
\centering
\includegraphics[angle=0,width=\columnwidth,keepaspectratio]{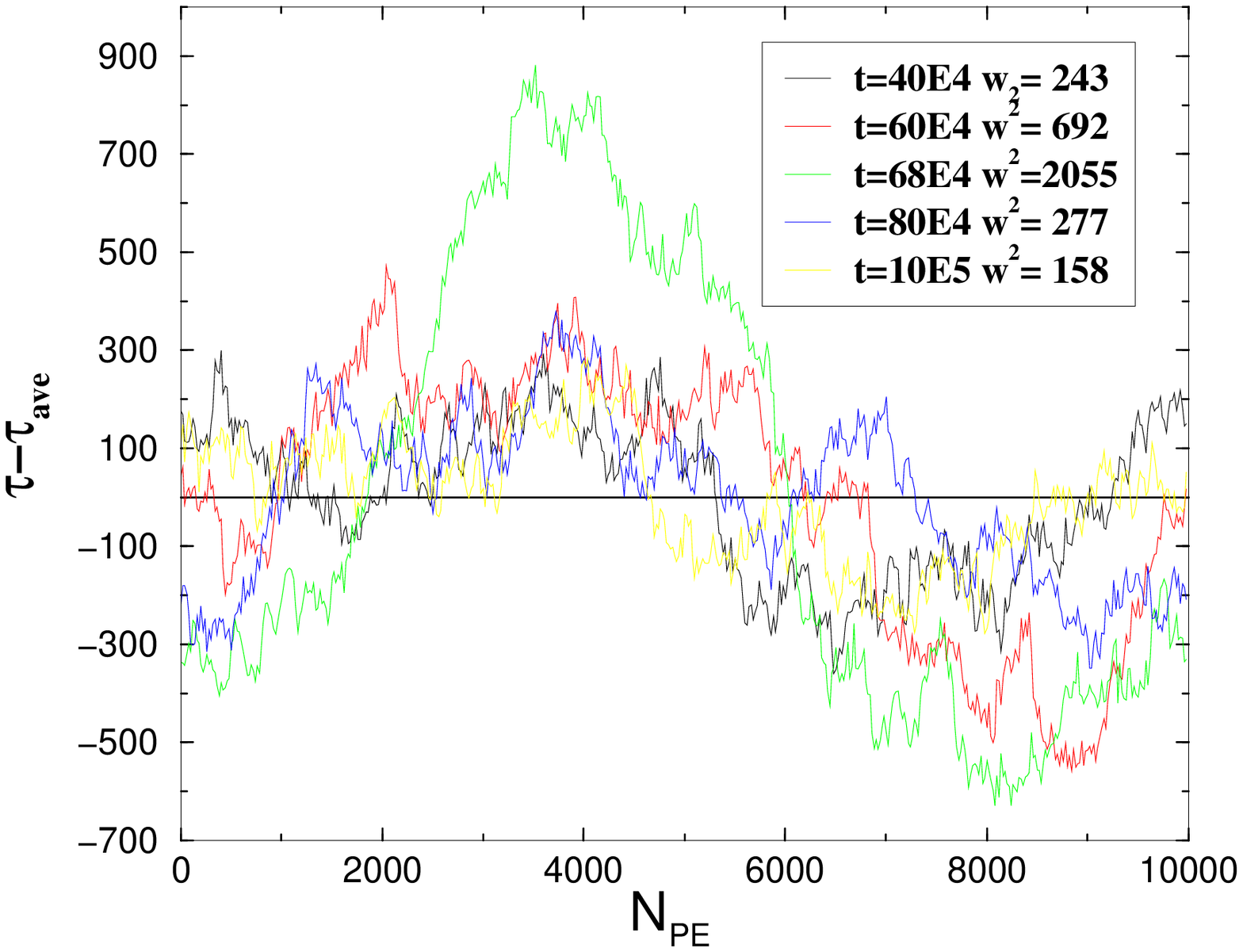}
\caption{Time horizon profiles at the fixed moments of time for the 
conservative PDES and $N_{\rm PE}=L=10^4$ for run number 4 of 
Fig.~\ref{NPE01000runs}.} 
\label{largfluc010000} 
\end{figure}

It is known, that the dynamics of solitons in the noisy Burgers
equation develops so-called Burgers
turbulence~\cite{Gurbatov,Chekhlov,Fog-noisy}. Despite the similarity
of the width evolution we have been unsuccessful in finding evidence
for Burgers turbulence. The tail of the width distribution (Figure~1a
of Korniss {\it et al.\/}) may be due to long-lived fluctuations,
rather than due to moving solitons.  Figure~\ref{NPE01000runs} shows
the evolution of the time-horizon width for some realizations of this
conservative PDES. A large fluctuation occurs in run number~4 at a
time $t\approx 68000$. The momentary picture of the time horizon
profile shown on Fig.~\ref{largfluc010000} looks like a soliton.
Nevertheless, a more detailed analysis is needed to claim that it is
indeed a soliton.  This evidence would be difficult to obtain, since
the noise in the PDES conservative iterative 
process~(\ref{horizon-Theta}) strongly masks the expected soliton-like
behavior.

The large fluctuation in the time horizon profile ($\approx 1600$)  is
comparable to the system size, $N_{\rm PE}=L=10^4$, for the case shown in
Fig.~\ref{largfluc010000}. In the next section, we will see that 
the iterative process~(\ref{horizon-Theta}) has steady-state solutions 
under some boundary conditions, and the time horizon profile shown in 
Fig.~\ref{largfluc010000} looks similar to it.

\section{Freeze-and-shift algorithm}
\label{Froze}

\subsection{Realizations of PDES.}

The conservative PDES algorithm of Korniss {\it et al.\/} 
is the ideal scheme, in
which each PE manages only one process and the LST  
is associated with the PE. We extend this to a different more 
general scheme than is present in other publications 
\cite{KornL00,Kola02,Kola03,Kola04}.

Consider the nodes to be vertices of a random graph. The links
between the vertices are associated with the possible communications, 
at least in one direction.  Imagine that we can group vertices in clusters, 
maximizing the connectivity within the cluster and minimizing the number of
links between clusters. Let us call those nodes without external links 
bulk nodes and the rest of the nodes surface nodes. Then, we can map this 
random graph onto the parallel computer architecture associating one cluster
of nodes with the one PE.

Practically, the nodes can be realized as threads~\cite{Threads}. Threads,
sometimes called lightweight processes, share the same local memory
with the other threads associated with the same PE.  
Hence, in a practical sense they do not require 
any extra communication between PEs to communicate with other threads on 
the same PE. Thus, the communication of the bulk nodes can
be organized in the most optimal way, using the conservative PDES 
implementation.  

We have to note that in our approach it is natural to associate the LST 
with nodes and not with the PEs, in contrast with previous work 
\cite{Korn00,KornL00,Kola02,Kola03,Kola04}. 
All nodes carry its own LST, even those 
belonging to the same PE.

The communication of surface nodes can be realized in a number of ways. 

First, surface nodes can communicate in a conservative manner. The LST
horizon will evolve as described by the Korniss {\it et al.\/}
scenario discussed in the previous section. Nevertheless, the
difference is that the average utilization is not given by $\langle
u\rangle$ (see expr.~(\ref{u-speed})).  Let $\ell$ be the number of
nodes per PE, so $L=N_{\rm PE}\ell$.  The probability of having a
chosen node not being at a local minimum of the LST is $1-\langle
u\rangle$.  Assuming complete randomness among the LST of the nodes,
{\it i.e.\/} using a mean-field type of argument \cite{Kola04} for
non-equilibrium properties of the time-evolving surface, gives that
the probability when choosing $\ell$ nodes that all of them are not at
a local minimum is $\left(1-\langle u\rangle\right)^\ell$.  Since a PE
can perform an update as long as any of its nodes are at a local
minimum, the mean-field argument gives the utilization to be
\begin{equation} \langle u\rangle = 1 - \left(1-\langle
u\rangle\right)^\ell \> . \label{Eq-u-node} \end{equation} Our
preliminary simulations show that the average utilization for 
$\ell=1$ depends on
the type of noise (all of which have mean noise per step of unity).  
For uniformly distributed noise it is $\left\langle
u\right\rangle_u\approx 0.267(4)$.  For Gaussian noise it is
$\left\langle u\right\rangle_{\rm G}\approx 0.258(5)$. These are to be
compared with Poisson noise with $\langle u\rangle =0.246410(7)$.  
Note that both Gaussian and uniformly distributed noise have an
average utilization larger than $1/4$.

In the second realization of PDES, the surface nodes postpone messages
within a discrete time window interval $t_{\rm w}$. The system would
evaluate in the conservative PDES manner the bulk nodes which lie at a
distance from the surface larger than $d \approx
\left(\ell-\sqrt{\ell^2-4t}\right)/2$ as later given by
Expr.~(\ref{EqRoottau}).  After the time $t_1 \approx \ell^2/4$ (see
Expr.~(\ref{Eqt1})) the freezing will reach the inner bulk nodes and
the evaluation will stop.  The system then could not propagate further
without an exchange of the messages between PEs to preserve
causality.  The whole system will at that time be ``frozen''.  The
second idea to implement this algorithm is the fast exchange of the
messages.  We propose to do that by redistributing nodes between PEs.
A simple realization will be discussed below.  We call this algorithm
the ``freeze-and-shift'' algorithm (FaS).  Note that the FaS algorithm
effectively separates the inter-processor communication from the
computations progressing on a PE.  Thus computer architectures that
are capable of simultaneously performing calculations and
inter-processor communication can be used effectively, even for
problems with fine-grained parallelism.  Another potential application
of the FaS algorithm is that it should allow simulations with
fine-grained parallelism to be performed on calculations on the grid.

The third possible realization is the optimistic scenario of PDES, in
which one assumes that all messages come in an order in which causality
is not violated. After a discrete time $t_{\rm w}$, one has to check
causality and send anti-messages to kill those which violate
causality. The process of generating anti-messages can lead to
``avalanches'' of the time-horizon.  These seem to finish with a time
horizon profile similar to the one generated by the FaS algorithm
introduced above. Clearly, the optimistic scenario is more time
consuming: first, the anti-message generation is not a short process;
and second, some substantial amount of work on the nodes processed will be
rejected. As described on a PE with $\ell$ nodes, the optimistic
scenario is some generalization of a combination of the first two
scenarios.  The understanding of the time horizon avalanche process in
the optimistic scenareo can be interesting by itself, but will not be
explored in this paper.

\subsection{Time horizon evolution for the FaS algorithm}

In this subsection, for reasons of clarity, we discuss the simplest
case of the FaS algorithm where the nodes effectively form a
one-dimensional graph. Despite this simplicity, the one-dimensional case
of the FaS algorithm can be applied to the parallel simulation of
one-dimensional partial differential equations (see
section~\ref{Disc}).  The one-dimensional FaS algorithm can also be
used in simulations of systems that have been organized into
disjoint spatial subsystem slices, with the slices forming a
one-dimensional graph.

Let us assume there are $\ell$ nodes (or threads) on each of $N_{\rm
PE}$ processor elements. For the case of Ising model simulations, each
node carries just one spin of the system of $L=\ell N_{\rm PE}$ spins.  
Nodes within a PE communicate in the conservative manner. The three
possible ways of the inter-PE communication correspond to different
{\em boundary conditions}: the conservative scenario is associated
with continuous boundary conditions;  the FaS scenario corresponds to
fixed boundary conditions; and the optimistic scenario - to free
boundary conditions (and associated LST roll-backs).

As we mentioned already, our implementation of conservative PDES is
analogous to that of the Korniss {\it et al\/} $\ell=1$ implementation
\cite{Korn00} and further extensions to larger $\ell$
\cite{Kola02,Kola03,Kola04}.  The only difference, and a major
difference for algorithm design, is that in our case the LST is
associated with the nodes and not with the PEs.

In the case of the FaS algorithm, the evolution of the LST horizon is quite 
different. 
\begin{figure}
\centering
\includegraphics[angle=0,width=\columnwidth,keepaspectratio]{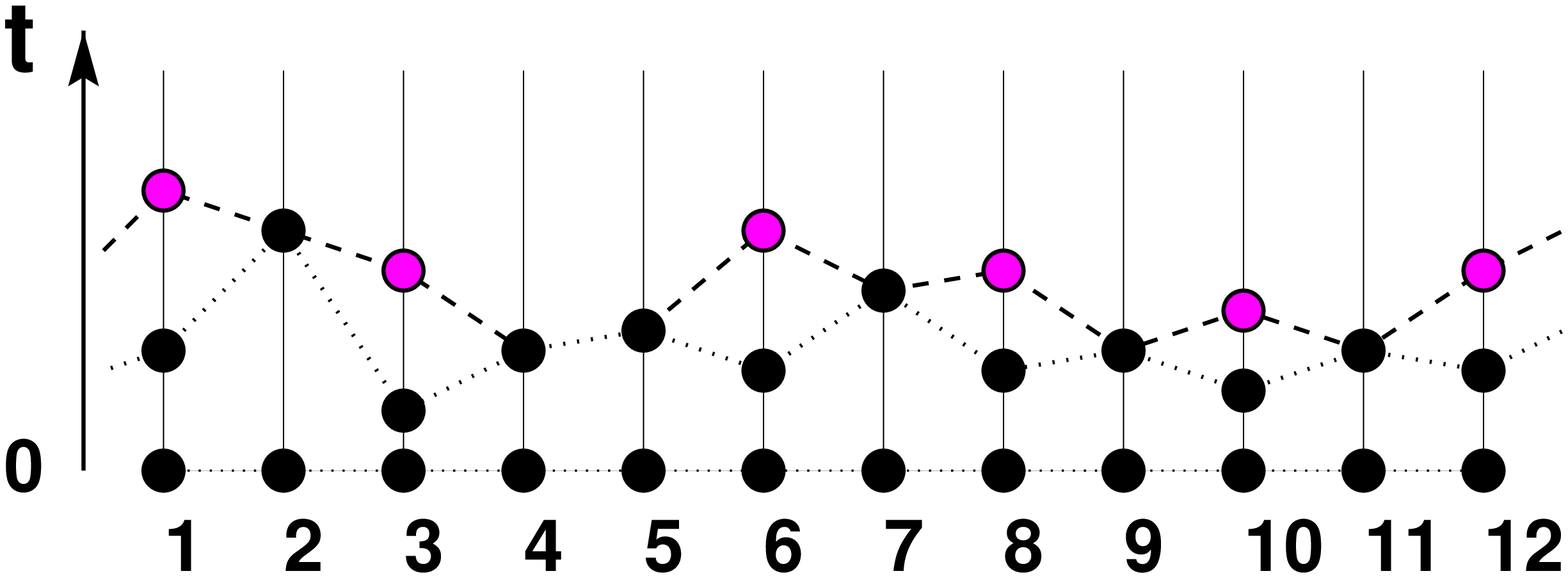}
\includegraphics[angle=0,width=\columnwidth,keepaspectratio]{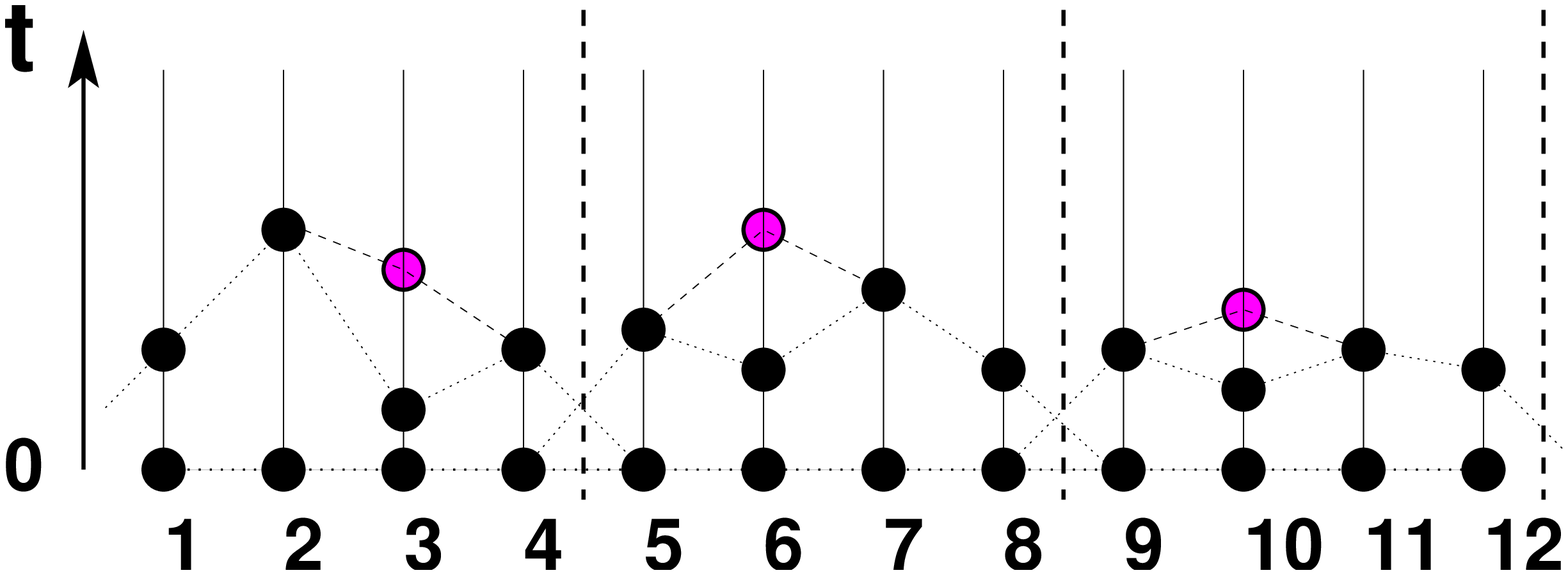}
\caption{Possible realizations of the first two steps of the time horizon 
evaluation: Conservative 
algorithm with $N_{\rm PE}=L=12$ and $\ell=1$ (Top) and the 
Freeze-and-Shift (FaS) Algorithm with $N_{\rm PE}=3$ with $\ell=4$ nodes 
each ($L=N_{\rm PE}\ell=12$).}
\label{pdes1-2}
\end{figure}

\begin{figure}
\centering
\includegraphics[angle=0,width=\columnwidth,keepaspectratio]{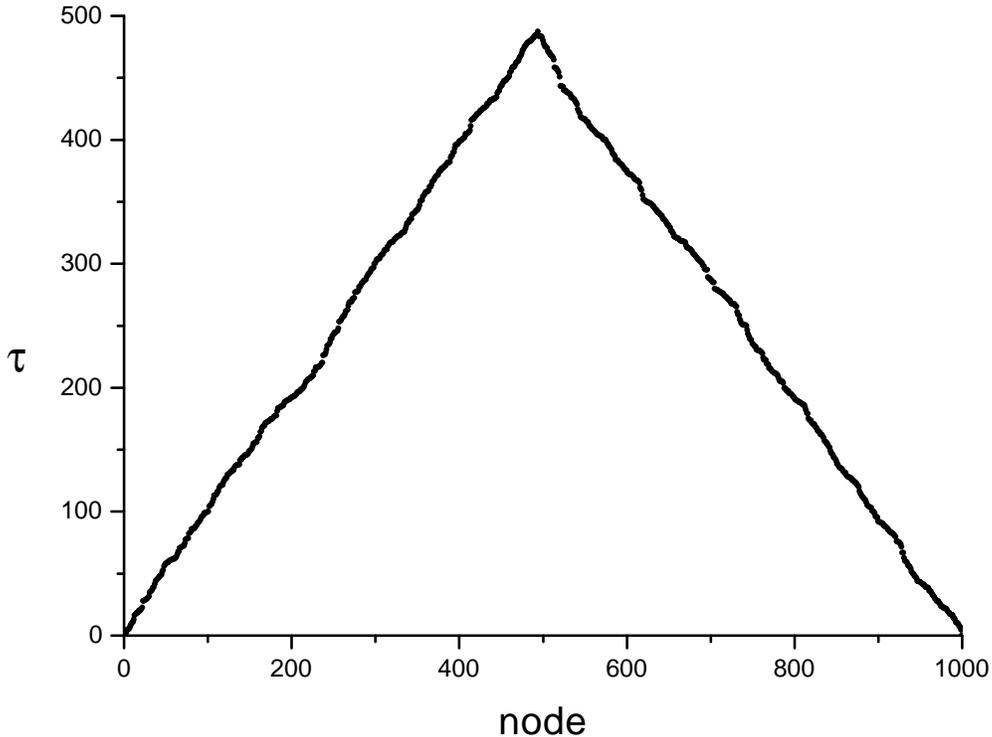}
\caption{Profile of the LST of $\ell=1000$ nodes with fixed 
boundaries at the leftmost and rightmost positions.}
\label{frozen-profile}
\end{figure}

In Fig.~\ref{pdes1-2} the two first steps of the possible evolution of 
the time horizon are sketched both for the 
case of the conservative scenario with $\ell=1$ and for the 
freeze-and-shift (FaS) scenario with $\ell=4$. 
Initially, at the time $t=0$ all $\tau_i=0$. In the first step
of PDES all the nodes proceed --- 
the condition~(\ref{horizon-eval}) is fulfilled for all nodes. 
The difference between the two scenarious starts just after the
first event (time step). In the top of Fig.~\ref{pdes1-2}, all the local
minima will advance in the conservative scenario. 
For the frozen boundaries of the FaS algorithm, 
the left and right boundaries on each PE are fixed, so these minima
cannot proceed.  
As the FaS algorithm progresses, the LST starts to develop a slope with 
an average angle $\psi$.  
The average value of this angle depends only on the mean of the noise
for a single time slice, with it's tangent being equal to that mean.  
This is because proceeding from a frozen node the question is: given that 
the next node can exceed the value of $\tau$ on the frozen node in the 
next time step what is the average distance at which its value of 
$\tau$ freezes ahead of the frozen node's $\tau$? 
Since we are using a mean noise of unity, the 
average angle will be $\psi=\pi/4$ for all types of noise.  This 
is seen in Fig.~\ref{frozen-profile}.  
Note that this profile would also develop if the algorithm always 
added unity to each updated node, as would be done in 
simulations of partial differential equations (see later on).  

The process of growth will stop at the moment the time
horizon reaches the hill knap.  
This average time can be calculated by assuming 
that it is given by the same time as the case where each node update 
advances $\tau$ by one unit.  This gives that 
\begin{equation}
t_1 = 2 \sum_{i=1}^{\ell/2} i \approx \frac{\ell^2}{4} \> .
\label{Eqt1}
\end{equation}
At that time the LST of each individual PE will look like a 
saw-tooth (Fig.~\ref{frozen-profile}).  
At time $t_1$ the entire time horizon will look like 
a saw with $N_{\rm PE}$ teeth. It is very important to realize that 
up to that
time all PEs were with high probability busy with an efficiency
of one~\footnote{We have ignored the extra book-keeping required for a 
single PE to decide which of its $\ell$ nodes can still be updated.  
This will often be the case in actual implementations.  However, if this 
is not the case one expects only a constant difference between 
serial execution and parallel simulations, with the constant independent 
of $N_{\rm PE}$}.

After a time $t_1$ the profile of the time horizon stops developing.
Thus the PE can on average perform $t_1$ node updates before 
the LST profile freezes.  
Figure~\ref{frozen-profile} shows a realization of this frozen steady-state
profile. The derivative of the profile $\phi_i=\tau_{i+1}-\tau_i$ will
represent a kink, the soliton-antisoliton pair. Contrary to the Burgers
equation~\cite{Fog-noisy}, these kinks are not moving but are 
in the steady-state. We
have an interesting result: by fixing the boundaries we select the
soliton-antisoliton solution of equation~(\ref{horizon-Theta}).

The mean value of the LST-horizon for $t\le t_1$ can also be 
calculated.  Consider the non-random case, starting with a 
flat distribution for $\tau=0$.  
The time for the LST-horizon to reach a plateau with all 
middle nodes at the same value of $\tau$ is
\begin{equation}
t = \sum_{i=1}^\tau \left(\ell-\tau\right)  \approx 
\tau\ell - \tau^2 \> .
\label{Eq_tau_t1}
\end{equation}
This equation can be solved for $\tau$ using the quadratic 
equation and choosing the physical root since $\tau\le\ell/2$, 
giving
\begin{equation}
\tau = \left(\ell-\sqrt{\ell^2-4t}\right)/2 \> .
\label{EqRoottau}
\end{equation}
For this configuration the average value of the LST is 
\begin{equation}
\langle\tau\rangle = 
\frac{\tau(\ell-\tau)}{\ell} \> .
\label{Eq_tau_t1_ave}
\end{equation}
Then substituting for $\tau$ from Eq.~(\ref{EqRoottau}) gives an 
expression for the time dependence for $\langle \tau\rangle$ 
for times $0\le t\le t_1$.  In a similar fashion an expression for 
the time evolution of $\langle w^2\rangle$ can be derived.  

At time $t_1$ the LST-horizon is frozen, and the question is 
how to proceed further with the simulation.  
The solution we propose is to
redistribute the nodes between the PEs. For example, let us shift
them by $\ell/2$ to the right (or left), so the tops of the hills
(saw-teeth) will be at the boundaries of the processors and fixed for the
next time window processing.

\begin{figure}
\centering
\includegraphics[angle=0,width=\columnwidth,keepaspectratio]{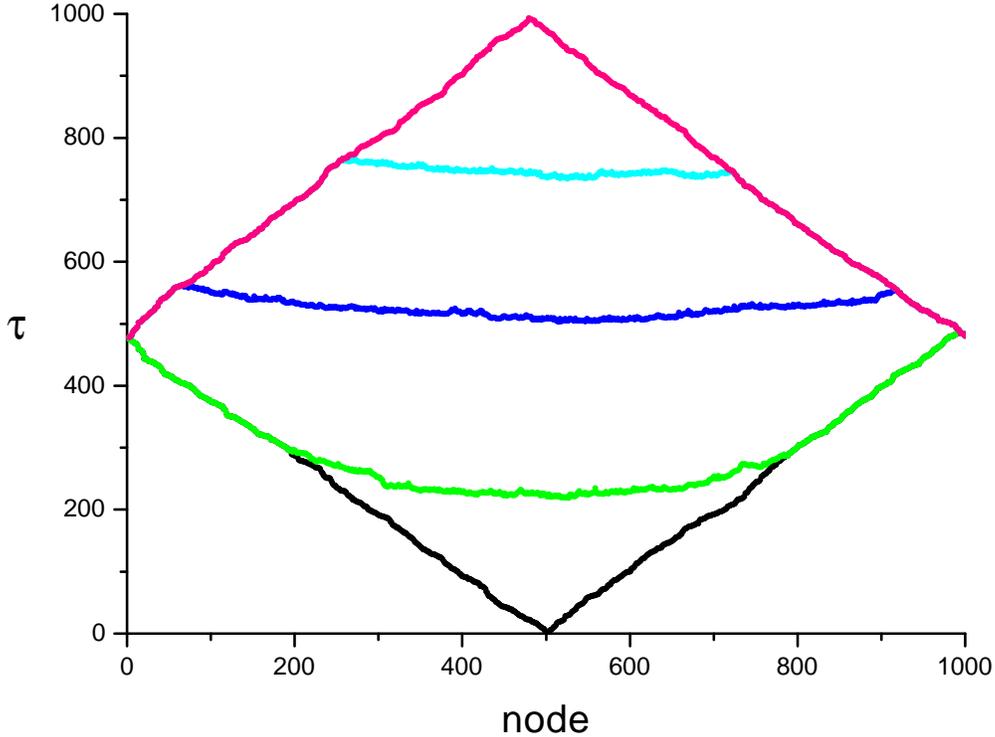}
\caption{Profile of the LST of $\ell=1000$ nodes at the time
$t_1$ after shifting by half (bottom line) and at
the time values $\approx 1.4 t_1$, $\approx 1.8 t_1$, $\approx 2.5 
t_1$, and $3 t_1$ from the second bottom to the top.}
\label{s-sh003}
\end{figure}

Futher evolution of the time horizon is illustrated in 
Fig.~\ref{s-sh003}. The lowest curve is the LST-horizon 
depicted on Fig.~\ref{frozen-profile} and cyclically shifted by $\ell/2$.
The LST-horizon will grow on average until the time 
$t_2\approx 3 t_1 = 3\ell^2/4$ 
at which time the next steady-state frozen solution will be
reached (the highest curve in Fig.~\ref{s-sh003}).

\begin{figure}
\centering
\includegraphics[angle=0,width=\columnwidth,keepaspectratio]{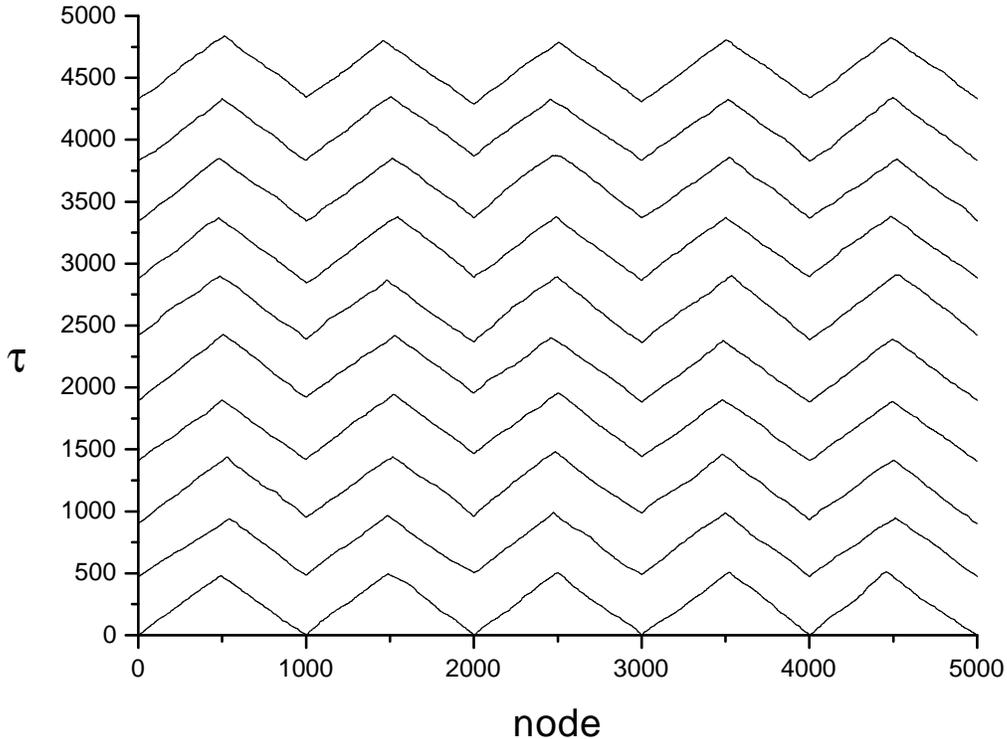}
\caption{Profile of a LST with $\ell=1000$ nodes in each of the 
$N_{\rm PE}=5$ PEs.}
\label{froze-shift}
\end{figure}

The process can be repeated by alternating freezing of the boundaries 
for a time window interval not longer than $2 t_1$ 
and shifting nodes
between PEs by $\ell/2$ for each time window.
Figure~\ref{froze-shift} illustrates this process for $\ell=1000$ nodes
and $N_{\rm PE}=5$ for ten shift-cycles with 
the time window of $2 t_1$.  Consequently there are about 
$20 t_1$ time steps (discrete events).  

\begin{figure}
\centering
\includegraphics[angle=270,width=\columnwidth,keepaspectratio]{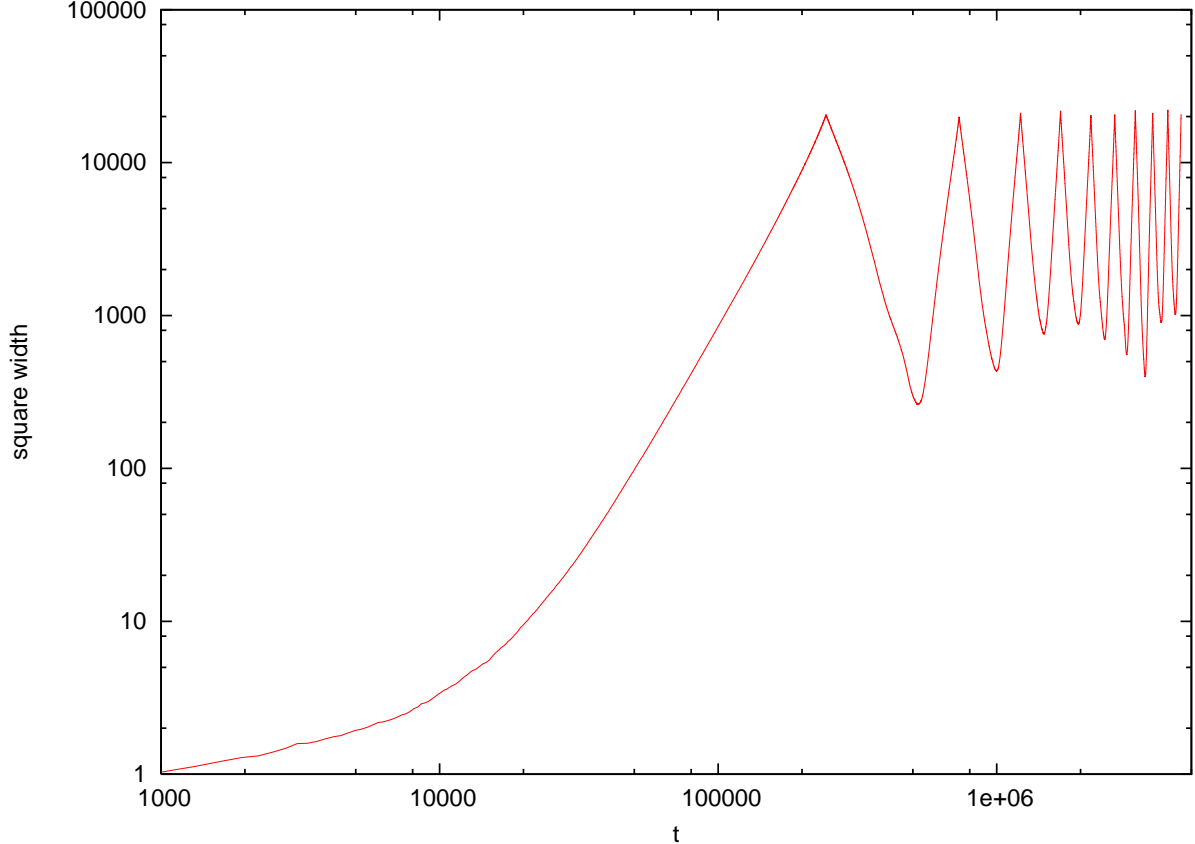}
\caption{The square of the LST width of the process depicted in 
Fig.~\ref{froze-shift}.}
\label{w-sh004}
\end{figure}

The square width $w^2(t)$ of the time horizon (LST) 
seems to evolve periodically
with time between a minimum and a maximum value.  
The evidence for this is seen in Fig.~\ref{w-sh004} in which
the LST evolution of $\left\langle w^2\right\rangle$ 
for $\ell=10^3$ and $N_{\rm PE}=5$ is shown. 
For the time between $t=2\times 10^4$ and up to the $t=2\times 10^5$ the 
LST-horizon width grows with the effective exponent $z=3/2$, it then 
stops at the shift moment, and oscillations start. 
This exponent is not associated with any stochastic process, but 
reflects the deterministic process of the saw-tooth's formation 
according with Expressions~(\ref{Eqt1}-\ref{Eq_tau_t1_ave}).
Probably, the maximum values of the LST width would follow 
the KPZ exponent, and this is a fruitful subject for future 
research.  
Figure~\ref{w-sh004} seems to show that in the 
FaS algorithm one can limit 
(effectively ``saturate'') the size of the surface width.  
Proven ways of saturating the surface width in conservative PDES 
simulations include imposing a fixed constraint on the width \cite{Kola02} 
and imposing small-world connections between PEs \cite{Korn03}.  
Although the freeze-and-shift method damps out the surface width, 
much larger studies would be required to see whether or not the 
governing universality class of the interface is still the 
KPZ universality class and the FaS algorithm just leads to a 
coarse-grained length in the KPZ equation.  

\begin{figure}
\centering
\includegraphics[angle=0,width=\columnwidth,keepaspectratio]{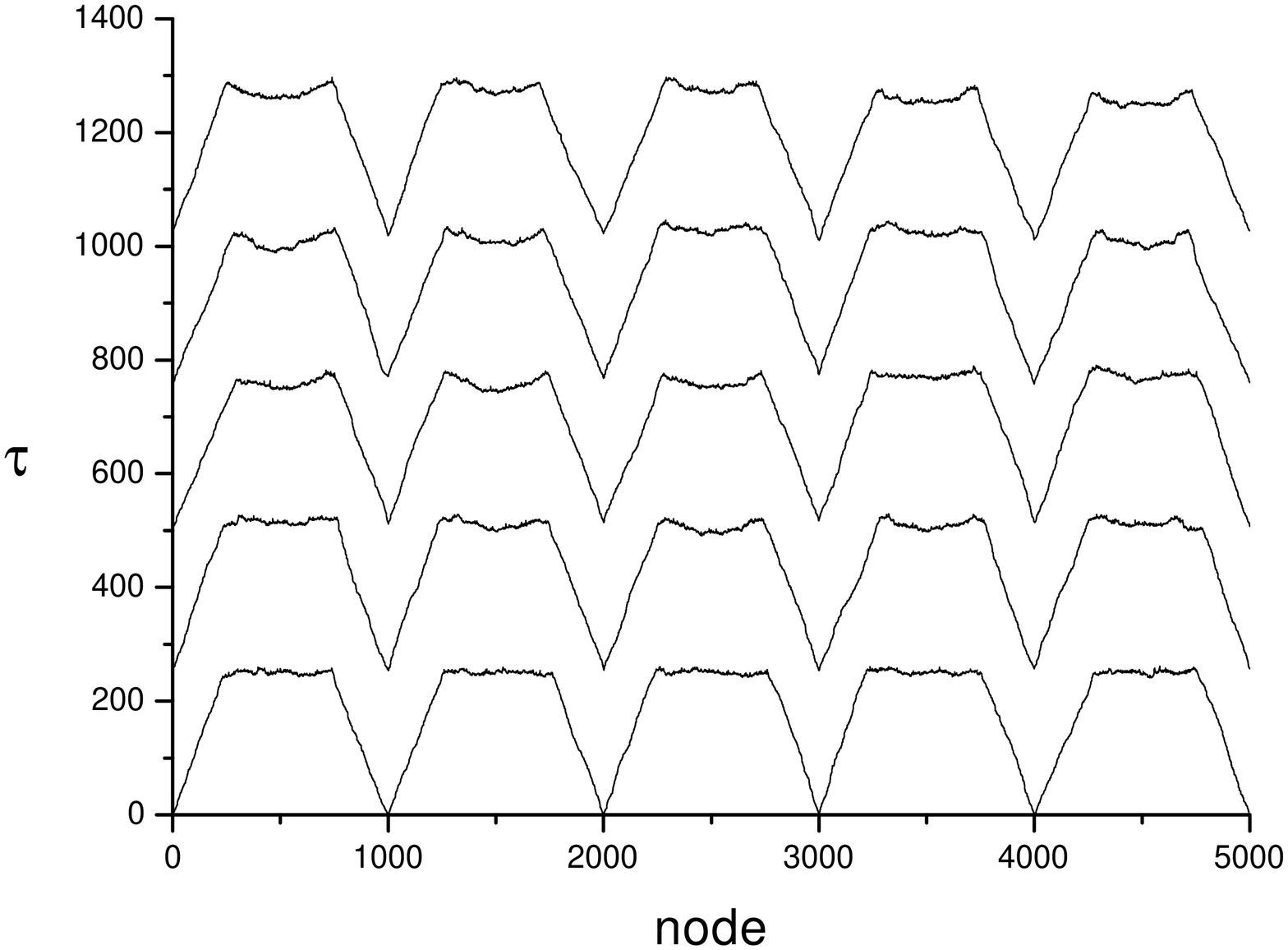}
\caption{Profile of the LST for $\ell=1000$ nodes on each of 
$N_{\rm PE}=5$ PEs. 
Curves from bottom to top are the profiles for $t_{\rm w}=\ell$ 
discrete events at the time 
moments $t_k=(2k-1) \ell$, $2k-1=1,3,5,7,9$.}
\label{s-sh005-odd}
\end{figure}

\begin{figure}
\centering
\includegraphics[angle=0,width=\columnwidth,keepaspectratio]{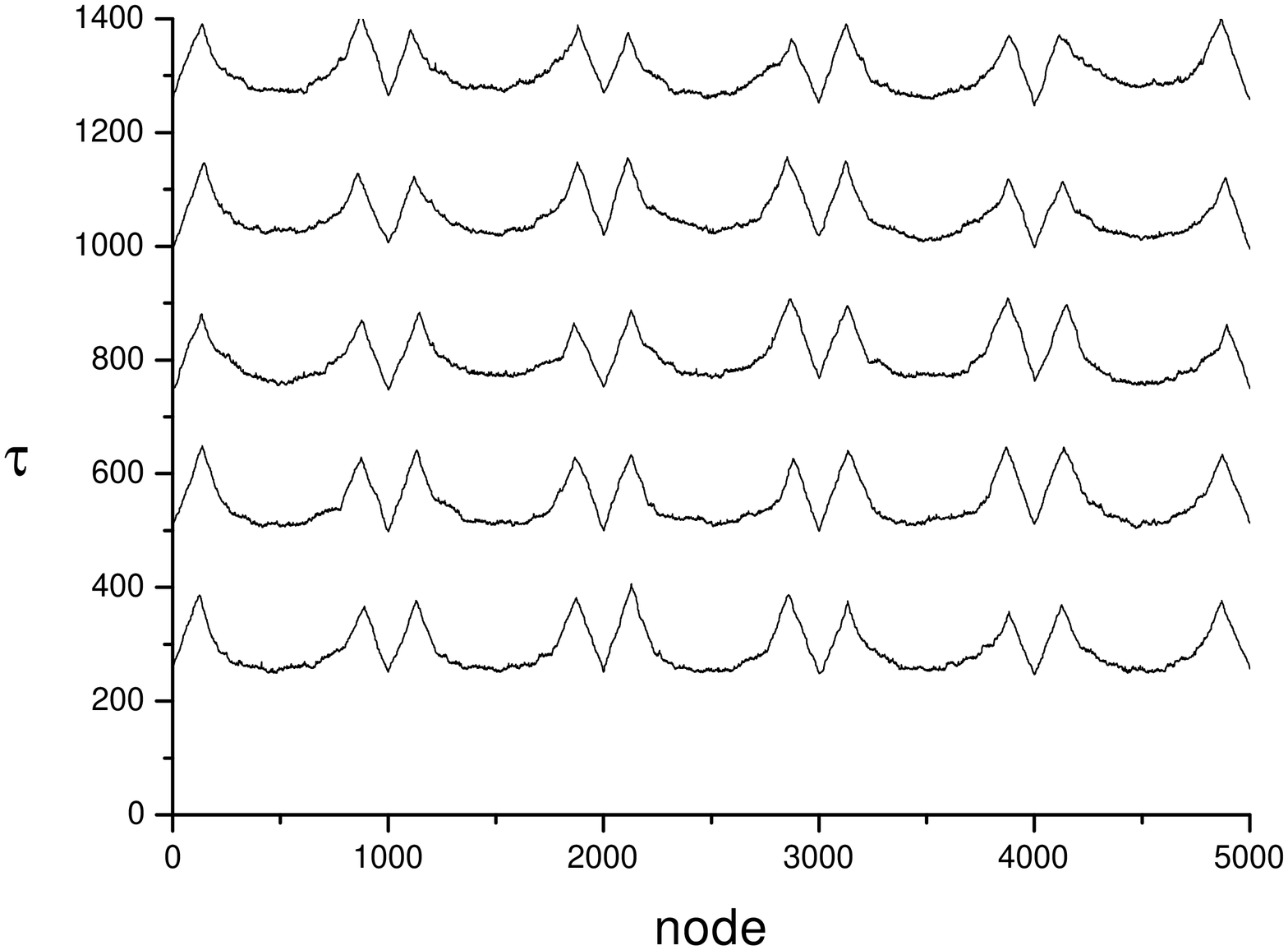}
\caption{Profile of the LST of $\ell=1000$ nodes in each of the 
$N_{\rm PE}=5$ PEs. 
Curves from bottom to top are the profiles for $t_{\rm w}=\ell$ 
discrete events at the times 
$t_k=2k \ell$, $2k=2,4,6,8,10$.}
\label{s-sh005-even}
\end{figure}

The time window $t_{\rm w}$ for the 
freeze-and-shift algorithm can be chosen as any
value in the interval $1 < t_{\rm w} \le t_1$.  
We can choose them shorter then
$t_1$, for example equal to $t_1/2$ as shown in
Fig.~\ref{s-sh005-odd} and Fig.~\ref{s-sh005-even} for the 
even and odd shifts, respectively. 
The difference between the maximum and minimum possible values of the 
LST will be smaller than in the case of larger $t_{\rm w}$.

\subsection{FaS on a two-dimensional lattice}
\label{2d}

The next interesting realization of the freeze-and-shift algorithm is the 
application to a two-dimensional model. 
Suppose we have $N_{\rm PE}$ each of which will simulate a 
$\ell\times\ell$ block of nodes (spins in kinetic Monte Carlo for the 
Ising model).  The total system size is $L^2=N_{\rm PE}\ell^2$ where 
$N_{\rm PE}$ should be a perfect square.  
We associate $\ell^2$ sites (spins) with $\ell^2$ threads running on each of 
the $N_{\rm PE}$ PEs. The evolution of the LST horizon is described by the 
two-dimensional iterative process
(compare with expression (\ref{horizon-Theta}))
\begin{eqnarray}
\tau_{i,j}(t+1)=
&&\tau_{i,j}(t)+\Theta\left(\tau_{i-1,j}(t)-\tau_{i,j}(t)\right)
\nonumber \\
&& \times \Theta\left(\tau_{i+1,j}(t)-\tau_{i,j}(t)\right)
\nonumber \\
&& \times \Theta\left(\tau_{i,j-1}(t)-\tau_{i,j}(t)\right)
\nonumber \\
&& \times \Theta\left(\tau_{i,j+1}(t)-\tau_{i,j}(t)\right) \eta_i(t)
\label{2d-horizon}
\end{eqnarray}
with the product of four theta-functions at the right hand side
and $(i,j)$ are coordinates of the lattice edges.  
For $\ell=1$ one expects that the average speed of time horizon for
the conservative PDES scenario will be approximately
$\left\langle u_2\right\rangle\approx 1/8$, 
two times slower than in one-dimensional
case.  Our computer simulation gives
$\left\langle u_2\right\rangle=0.1205(2)$ 
for a Poisson distribution of PDES arrival
times with unit mean value.  This value again depends on the 
distribution law of the random numbers used.

\begin{figure}
\centering
\includegraphics[angle=270,width=\columnwidth,keepaspectratio]{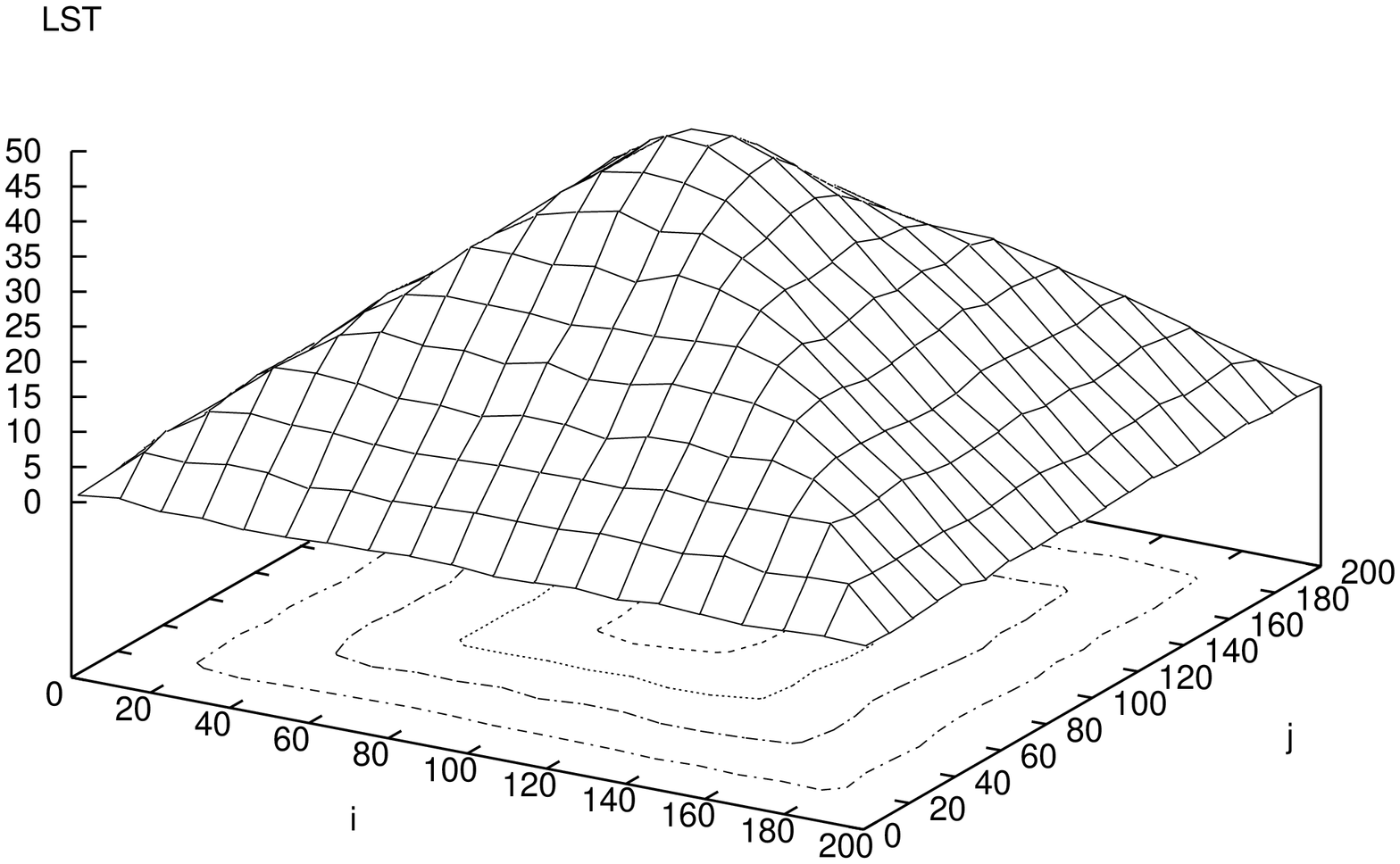}
\caption{Steady-state profile of the LST on a two-dimensional 
square lattice with linear number of nodes $\ell=200$.}
\label{pyramid}
\end{figure}

Figure~\ref{pyramid} shows the steady-state solution of process
(\ref{2d-horizon}) which looks like a pyramid with the slope $1/2$. 
This solution is achieved at a time given by the 
volume of the pyramid, 
$t_{\rm 2-sst} \approx \ell^3/12$.

We have to note that up to the time $t_{\rm 2-sst}$ the
efficiency of the FaS algorithm is practically unity. Shifting the 
nodes between the PEs can now be accomplished in many different ways.  
Shifting the nodes by $\ell/2$ in both directions of the lattice is 
equivalent to the sending of postponed messages in the language of PDES. 
Then, the LST frozen profile
will take a time of double $t_{\rm 2-sst}$ before reaching the second
frozen steady-state position.  Repeating this process of freezing and 
shifting, we can evolve our nodes in parallel as far as we wish.  
As in the one-dimensional FaS algorithm, the time between shifting 
can be any value given by $0<t_{\rm w}\le t_{\rm 2-sst}$.  
Note that this shifting can usually be implemented very effectively on modern 
computers which have fast methods of copying whole blocks of memory 
between processors, and may have the ability to simultaneously 
perform calculations and message passing.  

\subsection{FaS, partial differential equations, grid computing}
\label{pde}

The discussed algorithms can be effectively applied also to numerical
solutions for solving partial differential equations. 
In this case one has to solve
iteratively finite-differences equations defined on the lattice.  For
simplicity we will discuss partial differential equations of second
order in the one-dimensional case.  The common form can be written as
\begin{equation}
\psi_i(m)=F(\psi_i(m),\psi_{i-1}(m),\psi_{i+1}(m),
\psi_i(m-1),\psi_{i-1}(m-1),\psi_{i+1}(m-1);\Delta,\delta),
\label{pde-dif}
\end{equation}
where the index $i$ is associated with the space variable, the space
increment is $\delta$, and $m$ is associated with the time variable for 
which the increment is $\Delta$. 

We have to divide the entire one-dimensional lattice into 
$N_{\rm PE}$ pieces, each with 
$\ell$ lattice nodes, and then 
associate each piece with a single PE. 
The LST time increment is equal to $\Delta$, and is not
random in this case.  We freeze the nodes at the boundaries for each PE.  
The propagation of the algorithm on each PE 
will be the normal one, except the nodes that do not know the 
values of their own or neighboring nodes at both times $m$ and $m+\Delta$
will be frozen.  
This condition of the frozen boundaries between
neighboring PEs will form a LST horizon on each PE.  
For each time step $\Delta$ the space coordinates of the 
LST will be $\delta$.  
If the calculation of the right-hand side of Eq.~\ref{pde-dif} 
needs a large time compared with 
memory shifting between PEs, this algorithm could be very effectively 
implemented on a parallel computer architecture.

The scheme can be generalized to large dimensions of the lattice, i.e. 
to the many-dimensional partial differential equations, in the manner 
demonstrated in the previous subsection for FaS PDES.

Grid computing should enable geographically distributed heterogeneous 
computations to be performed if appropriate algorithms are available.  
For previous implementations with fine-grained parallelism, such 
algorithms are not available.  For example, for conservative PDES simulations 
implemented with one virtual time per PE, the calculation on a particular 
PE halts at irregular times (whenever the algorithm hits the surface 
node of the PE).  However, conservative PDES implemented with one 
virtual time per PE allows for an approximately regular and calculable 
number of time steps $t$ that a PE can perform before it needs to halt to 
preserve causality.  Furthermore, the FaS algorithm allows a decoupling of 
calculation and the communication.  Both of these facts can be important 
for grid computing.  One example is since grid communication paths are 
used by many people the time for communication between grid computers is 
not constant, the communication between grid computers can take place 
without the calculations on a computer having to wait for another computer.  
Furthermore, the communication may be timed so it is performed when 
others are historically utilizing the communications network the least (say 
during nights or weekends).  Thus the FaS algorithm should allow grid 
computations to be performed for PDES simulations and for 
numerical solutions of partial differential equations.  

\section{Discussion}
\label{Disc}

We have proposed a new algorithm for parallel discrete event simulations 
(PDES) which
allows for an effective realization on the parallel computers, clusters of
computers, and in grid computing.  
We call this algorithm the FaS (Freeze-and-Shift) algorithm.  
It effectively decouples the computation phase and communication phase 
for these parallel computations.  This allows, for example, 
the programmer to utilize fast block-memory-transfer commands.  Furthermore, 
it should allow the efficient simultaneous execution 
of both computation and communication 
hardware when they are performed by separate hardware.  

There are several essential points of our FaS algorithm. First, we associate
a LST with the nodes rather then individual processing elements (PEs). 
Second, we group the nodes using some
rule (see later) and associate each group with a PE 
(usually the processor running the one process). The nodes are realized as
threads, which share the same memory within one process.  This allows for 
a fast realization of the conservative PDES rules within a PE. 
PEs do not communicate with other PEs for some time interval window 
$t_{\rm w}$ (the frozen part of the algorithm), 
and after that time the message exchange is realized as 
part-of-the-memory shifting between PEs (the shift part of the 
algorithm). The last part can be very efficiently realized on some 
parallel architectures.  The width of the LST-horizon, which characterizes 
the difference in the LST between different shifts is under control and 
depends on the number of nodes $\ell$ per PE. The FaS algorithm 
guaranties the efficiency of the simulations and idleness of PEs 
should be nearly equal to zero for balanced simulations.

\begin{figure}
\centering
\includegraphics[angle=0,width=\columnwidth,keepaspectratio]{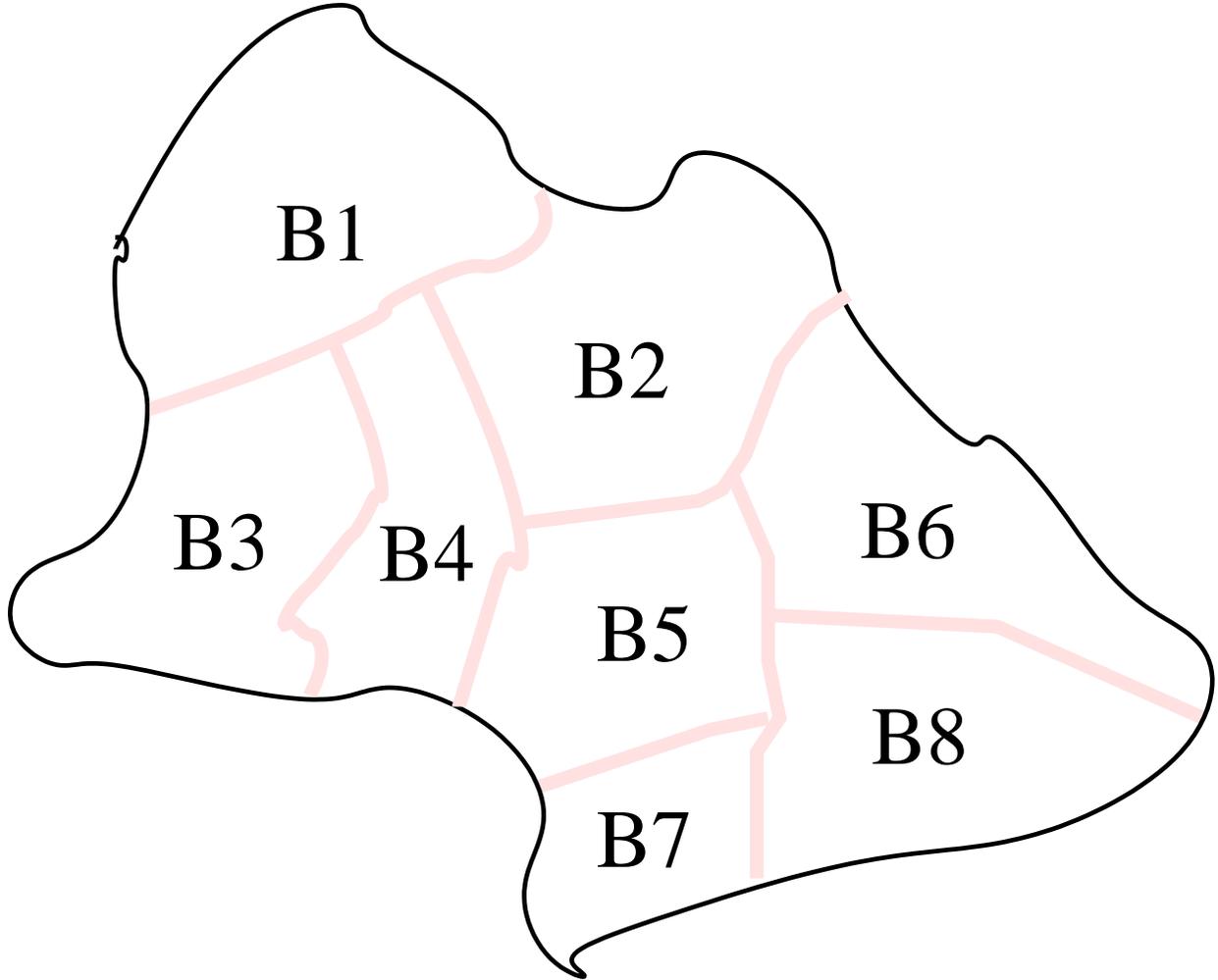}
\caption{Nodes manifold partitioning into eight groups of bulk nodes, 
each associated with a single PE during the frozen phase of the FaS 
algorithm.}
\label{general}
\end{figure}

The general scheme can be sketched as in Fig.~\ref{general}, where we have 
partitioned the nodes into groups (associated with PEs during the 
freeze part of the FaS algorithm) and we freeze development of those nodes 
within the interfaces.  
The efficiency of the algorithm depends on the smallness of the ratio of
the interface area to the bulk area, and on the new (alternating) interface
area we have to create by the shifting of nodes between PEs (the 
alternating partitioning of the node manifold).  

Analysis of several realizations demonstrate the efficiency of the general 
idea of the FaS algorithm.  Actual implementations on parallel computers, 
heterogeneous compute clusters, and computer grids 
for problems of interest will determine the ultimate 
usefulness of the FaS algorithm.

\acknowledgments

Supported in part by National Science Foundation grants 
DMR-0113049 and DMR-0120310, and by Russian Foundation
for Basic Research.  
LNS thanks Department of Physics and Astronomy and the 
ERC Center for Computational Sciences 
at Mississippi State University for their hospitality.


\begin{thebibliography}{12345678901234}
\frenchspacing

\bibitem{Fuji90} R.M.\ Fujimoto, Comm.\ ACM, {\bf 33} (1990) 31.

\bibitem{Jeff85} D.R.\ Jefferson, Assoc.\ Comput.\ Mach.\ Trans.\ 
Programming Languages and Systems {\bf 7}, (1985) 404.  

\bibitem{Over00} P.M.A.\ Sloot, B.J.\ Overeinder, and A.\ Schoneveld, 
Comp.\ Phys.\ Comm., {\bf 142} (2001) 76; 
B.J.\ Overeinder, PhD Thesis, Univ.\ of Amsterdam, 2000.

\bibitem{Korn99} G.\ Korniss, M.A.\ Novotny, and P.A.\ Rikvold, J.\ 
Comput.\ Phys., {\bf 153} (1999) 488.

\bibitem{Korn00} G.\ Korniss, Z.\ Toroczkai, M.A.\ Novotny, and 
P.A.\ Rikvold,  Phys.\ Rev.\ Lett., {\bf 84} (2000) 1351.

\bibitem{Kola02} A.\ Kolakowska, M.A.\ Novotny, and G.\ Korniss, 
Phys.\ Rev.\ E {\bf 67} (2002) 046703.  

\bibitem{Kola03} A.\ Kolakowska, M.A.\ Novotny, and P.A.\ Rikvold, 
Phys.\ Rev.\ E {\bf 68} (2003) 046705.  

\bibitem{Kola04} A.\ Kolakowska and M.A.\ Novotny, 
Phys.\ Rev.\ E, submitted, preprint cond-mat/0311015.  

\bibitem{Gurbatov} S.N.\ Gurbatov, A.N.\ Malakhov, A.I.\ Saichev, {\it
Nonlinear random waves and turbulence in nondispersive media:  waves,
rays, particles}, (Manchester University Press, Manchester, 1991)

\bibitem{Chekhlov} A.\ Chekhlov, and V.\ Yakhot, Phys.\ Rev.\ E {\bf 
51} (1995) R2739.

\bibitem{Fog-noisy} H.C.\ Fogedby, Phys.\ Rev.\ E., {\bf 57} (1998) 4943.

\bibitem{Threads} A.S.\ Tanenbaum, Modern Operation Systems. 
(Prentice Hall, Upper Saddle River, 2001). 

\bibitem{Luba88} B.D.\ Lubachevsky, Complex Syst., {\bf 1} (1987) 1099.; 
J.\ Comput.\ Phys., {\bf 75} (1988) 103.

\bibitem{KornL00} G.\ Korniss, M.A.\ Novotny, Z.\ Toroczkai, and 
P.A.\  Rikvold, in 
{\em Computer Simulation Studies in Condensed Matter Physics~XIII}, 
editors D.P.\ Landau, S.P.\ Lewis, and H.-B.\ Sch{\"u}ttler 
(Springer Verlag, Heidelberg, 2001), p.~183.  

\bibitem{Kard86} M.\ Kardar,  G.\ Parisi, and Y.-C.\ Zhang, 
Phys.\ Rev.\ Lett., {\bf 56} (1986) 889.

\bibitem{Korn03} G.\ Korniss, M.A.\ Novotny, H.\ Guclu, Z.\ Toroczkai,
and P.A.\ Rikvold, 
Science {\bf 299} (2003) 677.  

\end{thebibliography}
\end{document}